\documentclass[aps,nofootinbib,prd,superscriptaddress,tightenlines,notitlepage,twocolumn,showpacs]{revtex4}

\usepackage{natbib}
\usepackage[colorlinks]{hyperref}
\usepackage{tabularx}
\usepackage{graphicx}
\usepackage{amssymb,bm}
\usepackage{xcolor}
\usepackage{amsmath}
\usepackage[varg]{txfonts}
\usepackage{enumerate}
\usepackage[normalem]{ulem}

\newcommand{\apjl}{Astrophys. J. Lett.}
\newcommand{\apjs}{Astrophys. J. Suppl. Ser.}
\newcommand{\aap}{Astron. \& Astrophys.}

\newcommand{\mnras}{Mon. Not. R. Astron. Soc.}

\newcommand{\PMO}{\affiliation{Purple Mountain Observatory, Chinese Academy of Sciences, Nanjing 210034, China}}

\newcommand{\GXU}{\affiliation{Guangxi Key Laboratory for Relativistic Astrophysics, Guangxi University, Nanning 530004, China}}

\begin{document}

\title{Precision Test of the Weak Equivalence Principle from Gamma-Ray Burst Polarization}
\author{Jun-Jie Wei}\thanks{jjwei@pmo.ac.cn}\PMO\GXU
\author{Xue-Feng Wu}\thanks{xfwu@pmo.ac.cn}\PMO

\begin{abstract}
If the weak equivalence principle (WEP) is broken, the measured values of the parametrized post-Newtonian parameter $\gamma$
from photons with left- and right-handed circular polarizations should differ slightly, leading to the arrival-time difference of
these two circular components. Thus, the polarization vector of a linearly polarized light may rotate during the propagation.
The rotation angle of the polarization vector depends on both the photon energy and the distance of the source.
It is believed that if the rotation angle differs by more than $\pi/2$ over an energy range, then the net polarization
of the signal would be significantly suppressed and could not be as high as the observed level. Thus, the
detection of highly polarized photons implies that the relative rotation angle ($\Delta\Theta$)
should not be too large. In this paper, we give a detailed calculation on the evolution of gamma-ray burst (GRB)
polarization arising from a possible violation of the WEP, and we find that more than $60\%$
of the initial polarization degree can be conserved even if $\Delta\Theta$ is larger than $\pi/2$.
In addition, to tightly constrain the WEP violation, GRBs with harder spectra and polarization observations
in a wider energy range seem to be favored. Applying our formulas to the measurements of linear polarization from GRB 110721A and
GRB 061122, we obtain strict limits on the differences of the $\gamma$ values as low as
$\Delta\gamma<1.3\times10^{-33}$ and $\Delta\gamma<0.8\times10^{-33}$. These provide the most stringent limits
to date on a deviation from the WEP, improving at least 6 orders of magnitude over previous bounds.
\end{abstract}

\pacs{04.80.Cc, 95.30.Sf, 98.70.Dk, 98.70.Rz}

\maketitle

\section{Introduction}
The weak equivalence principle (WEP) states that the trajectory of any freely falling, uncharged test body
is independent of its internal structure and composition \citep{2006LRR.....9....3W}.
In the parametrized post-Newtonian (PPN) formalism, the validity of the WEP can be characterized by the
same value of the PPN parameter $\gamma$ for different test particles \citep{2006LRR.....9....3W}.
According to the Shapiro time delay effect \citep{1964PhRvL..13..789S}, the time interval required
for particles to travel across a given distance is longer in the presence of a gravitational potential $U(r)$ by
\begin{equation}
t_{\rm gra}=-\frac{1+\gamma}{c^3}\int_{r_e}^{r_o}~U(r)dr\;,
\end{equation}
where $r_e$ and $r_o$ represent the locations of the emitting source and the observer, respectively.
Generally, the gravitational potential sourced by the matter density along the propagation path is considered.
All metric theories of gravity satisfying the WEP predict that any two different species of massless neutral particles,
or two of the same species of particles with different internal structures (e.g., energies or polarizations)
or different compositions, must follow identical trajectories and undergo the same Shapiro delay.
That is, all metric theories predict that the value of $\gamma$ should be the same
for all test particles.
Therefore, any two test particles, if emitted simultaneously from the same source and passing through
the same gravitational field with the same speed, should reach us at the same time.

The arrival time delays of different messenger particles have been applied to test the WEP accuracy
through the relative differential variations of the $\gamma$ values
\citep{1988PhRvL..60..173L,2016JCAP...08..031W,2015ApJ...810..121G}.
Since the polarization is considered to be part of the internal structure of photons,
\citet{2017PhRvD..95j3004W} proposed that polarization measurements from gamma-ray bursts (GRBs)
can also serve as an ideal test bed to probe the WEP.
For a linearly polarized light, it is a superposition of two monochromatic waves with opposite circular polarizations
(labeled with \emph{l} and \emph{r}).
Once the WEP fails, then different $\gamma$ values are measured with left- and right-handed circularly polarized photons
($\gamma_{l}\neq\gamma_{r}$), which leads to the slight arrival-time difference of these two circular components.
The relative time delay is then expressed as
\begin{equation}
\Delta t_{\rm gra}=\left|\frac{\Delta\gamma}{c^3}\int_{r_e}^{r_o}~U(r)dr\right|\;,
\label{eq:delta-tgra}
\end{equation}
where $\Delta\gamma=\gamma_{l}-\gamma_{r}$ is the difference of the $\gamma$ values for different
circular polarization states. This relative time delay would result in an energy-dependent rotation of
the polarization vector of a linear polarized light.
The WEP test can therefore be performed with observations of linear polarization.
It is believed that if the rotation angle differs by more than $\pi/2$ over a range of energies
$[E_{1},\;E_{2}]$ (where $E_{2}>E_{1}$), then the net polarization of the signal is severely suppressed and cannot be as high
as the observed level \citep{2012PhRvL.109x1104T}. Hence, the detection of high polarization implies that
the relative rotation angle ($\Delta\Theta$) should not be too large. Assuming that the arrival time delay
of two circular polarization states is caused by the Milky Way's gravitational potential,
and setting the upper limit of $\Delta\Theta$ to be $2\pi$, \citet{2017MNRAS.469L..36Y} obtained
the current best limit on a deviation from the WEP of $\Delta\gamma<1.6\times10^{-27}$ from the polarimetric
data of GRB 110721A.

However, this best WEP test was based on the assumption that the relative rotation angle $\Delta\Theta$ is
smaller than $2\pi$ \citep{2017MNRAS.469L..36Y}, one interesting question to ask is whether the net polarization
is significantly depleted when $\Delta\Theta$ is large. Here we give a detailed calculation on the evolution
of GRB polarization arising from a possible violation of the WEP as a function of $\Delta\Theta$, and
we show that a considerable amount of polarization can be conserved even if $\Delta\Theta$ is large. We then employ our general formulas to the GRB
polarimetric data and thereby obtain the hitherto most stringent limit on the WEP violation, improving previous results
by 6 orders of magnitude.

\section{General Formulas}
\subsection{The rotation angle of linear polarization}
A possible violation of the WEP can lead to the arrival-time difference of photons
with left- and right-handed circular polarizations, which results in an energy-dependent rotation of the polarization vector
of linear polarized photons.
With the calculated relative time delay $\Delta t_{\rm gra}$, the rotation angle during
the propagation from the source at redshift $z$ to the observer is given by \citep{2017MNRAS.469L..36Y}
\begin{equation}
\Delta\theta\left(E\right)=\Delta t_{\rm gra}(z)\frac{2\pi c}{\lambda}=\Delta t_{\rm gra}(z)\frac{E}{\hbar}\;,
\label{eq:delta-theta}
\end{equation}
where $E$ is the observed energy.

To estimate $\Delta t_{\rm gra}$ with Eq.~(\ref{eq:delta-tgra}), one has to figure out the gravitational
potential $U(r)$ along the propagation path. For cosmic sources, the Shapiro delay arising from
the gravitational potential of the large-scale structure
has been confirmed to be more important than the Milky Way's and
the source host galaxy's gravity \citep{2016JHEAp...9...35L}.
As the closest and most massive gravitational body to our Milky Way, the Laniakea supercluster of galaxies
\citep{2014Natur.513...71T} was usually adopted as the deflector in the Shapiro delay tests
\cite{2016JCAP...08..031W,2017PhRvD..95j3004W}.
As long as the distance of the source is far beyond the scale of Laniakea, Laniakea can be treated
as a point-mass approximation when calculating the gravitational potential.
Adopting a Keplerian potential $U(r)=-GM/R$ for Laniakea, one then has \citep{1988PhRvL..60..173L}
\begin{equation}\label{eq:gammadiff}
\Delta t_{\rm gra}= \Delta\gamma \frac{GM_{\rm L}}{c^{3}}
\ln \left\{ \frac{ \left[d+\left(d^{2}-b^{2}\right)^{1/2}\right] \left[r_{L}+s_{\rm n}\left(r_{L}^{2}-b^{2}\right)^{1/2}\right] }{b^{2}} \right\}\;,
\end{equation}
where $M_{\rm L}\simeq10^{17}M_{\odot}$ is the Laniakea mass \citep{2014Natur.513...71T}, $d$ is the approximate distance
from the source to the observer, $b$ denotes the impact parameter of the light path relative to the Laniakea center,
$r_{L}\simeq77$ Mpc represents the distance from the Laniakea center to the observer \citep{1988ApJ...326...19L},
and $s_{\rm n}=\pm1$ is the sign of the correction of the source direction, where $s_{\rm n}=+1$ ($s_{\rm n}=-1$)
corresponds to the source located along the direction of the Laniakea (anti-Laniakea) center.

\subsection{The polarization evolution of GRB photons}
Following \citet{2016MNRAS.463..375L}, we suppose that a beam of noncoherent light emits from
the source, in which the polarization direction is in the $xy$-plane and the propagation direction is chosen as
the $z$-axis. The intensity of photons within the infinitesimal azimuth angle interval $d\theta$ and
the infinitesimal energy interval $dE$ can be expressed as
\begin{equation}\label{eq:intensity1}
  dj(\theta,E)=j_0f(\theta)EN(E)d\theta dE\;,
\end{equation}
where $j_0$ is a normalized constant, $f(\theta)$ is a periodic function of $\theta$ with period $\pi$, and $N(E)$
is the photon spectrum. As photon intensity is proportional to the electric vector squared, the intensity projected
onto the direction of azimuth angle $\varphi$ is taken to be
\begin{equation}\label{eq:intensity2}
  dj_{\varphi}(\theta,E)=j_0f(\theta)EN(E)\cos^2(\varphi-\theta)d\theta dE\;.
\end{equation}
Thus, the total intensity of photons polarized along the direction $\varphi$ is given by
\begin{equation}\label{eq:intensity3}
  I(\varphi)=\int dj_{\varphi}(\theta,E)=\int_0^{\pi}d\theta\int_{E_1}^{E_2}dE~j_0f(\theta)EN(E)\cos^2(\varphi-\theta)\;,
\end{equation}
where $E_1<E<E_2$ is the energy range of the photon spectrum.
The polarization degree is defined as \citep{1979rpa..book.....R}
\begin{equation}\label{eq:polarization1}
  \Pi=\frac{I_{\rm max}-I_{\rm min}}{I_{\rm max}+I_{\rm min}}\;,
\end{equation}
where $I_{\rm min}$ and $I_{\rm max}$ are the minimum and maximum values of $I(\varphi)$, respectively.

It is obvious from Eq.~(\ref{eq:delta-theta}) that the rotation angle of the polarization vector has
an energy dependence. Let $\Delta\Theta\equiv\Delta\theta(E_2)-\Delta\theta(E_1)$ to be the difference of
rotation angles over an energy range of $E_1<E<E_2$; then Eq. (\ref{eq:delta-theta}) can be rephrased as
\begin{equation}\label{eq:delta-theta2}
  \Delta\theta(E)=\Delta\Theta\frac{E}{E_2-E_1}\;.
\end{equation}
By replacing $f(\theta)$ with $f(\theta+\Delta\theta(E))$, the received photon intensity can be derived from
Eq. (\ref{eq:intensity3}), i.e.,
\begin{equation}\label{eq:intensity4}
  I'(\varphi)=\int_0^{\pi}d\theta\int_{E_1}^{E_2}dE~j_0f(\theta+\Delta\theta(E))EN(E)\cos^2(\varphi-\theta).
\end{equation}
Accordingly, the observed polarization degree is
\begin{equation}\label{eq:polarization2}
  \Pi'=\frac{I'_{\rm max}-I'_{\rm min}}{I'_{\rm max}+I'_{\rm min}},
\end{equation}
where $I'_{\rm min}$ and $I'_{\rm max}$ are the minimum and maximum values of $I'(\varphi)$, respectively.

For most GRBs, the spectrum in a wide energy range of a few keV to tens MeV can be well described by
a broken power law, known as the Band function \citep{1993ApJ...413..281B}. Three independent spectral parameters are involved:
the low-energy photon index ($\alpha$), the high-energy photon index ($\beta$),
and the spectral peak energy ($E_p$). Their typical values are
$\alpha\approx-1.1$, $\beta\approx-2.2$, and $E_p\approx250$ keV, respectively \citep{2000ApJS..126...19P}.
But, the working energy band of current gamma-ray polarimeters is very narrow.
In such a narrow energy band, the spectrum can be adequately fit with a single power law,
$N(E)\propto E^{\Gamma}$. For simplicity, in the following we assume that the GRB spectrum
takes the form of a single power law with the index $\Gamma$ ranging from $-1.1$ to $-2.2$.

Here we consider that photons are initially completely polarized, e.g., along the $x$-axis.
To ensure $f(\theta)$ is periodic, i.e., $f(\theta+\pi)=f(\theta)$, the sum of $\delta$-functions is adopted \cite{2016MNRAS.463..375L}:
\begin{equation}\label{eq:f-theta}
  f(\theta)=\sum_{n=-\infty}^{+\infty}\delta(\theta-n\pi)\;.
\end{equation}
Substituting $N(E)=E^{\Gamma}$ and Eq. (\ref{eq:f-theta}) into Eq. (\ref{eq:intensity3}), we obtain
the initial photon intensity
\begin{equation}
  I(\varphi)=j_0\cos^2\varphi \int_{E_1}^{E_2}dE~E^{\Gamma+1}\;.
\end{equation}
It is easy to see that $I_{\rm max}=I(0)$ and $I_{\rm min}=I(\pi/2)=0$, which implies that the initial polarization degree
is $100\%$. Substituting $N(E)=E^{\Gamma}$ and Eqs. (\ref{eq:delta-theta2}) and (\ref{eq:f-theta})
into Eq. (\ref{eq:intensity4}), the observed photon intensity simplifies to
\begin{equation}
  I'(\varphi)=j_0 \int_{E_1}^{E_2}dE~E^{\Gamma+1}\cos^2\left(\varphi+\Delta\Theta\frac{E}{E_2-E_1}\right)\;.
\end{equation}

\begin{figure}
\vskip-0.3in
\centerline{\includegraphics[angle=0,width=1.0\hsize]{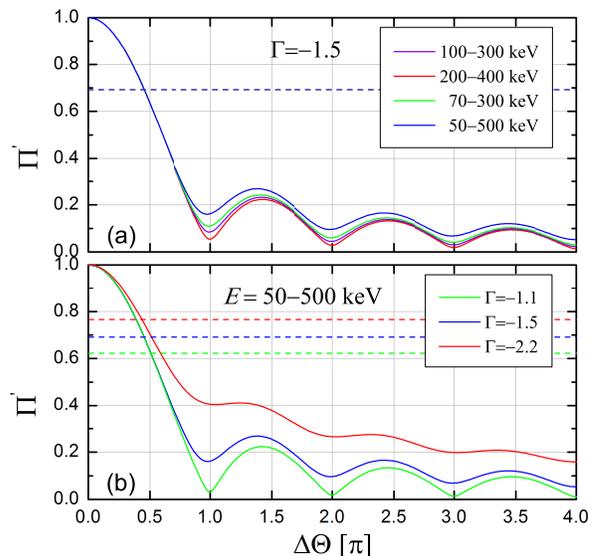}}
\vskip-0.4in
\caption{Polarization degree $\Pi'$ as a function of the relative rotation angle $\Delta\Theta$ (solid lines).
(a): The photon index $\Gamma$ is fixed at $-1.5$. (b): The energy band is fixed in 50--500 keV.
Dashed lines represent the level of polarization that is produced by synchrotron emission.}
\label{fig:f1}
\end{figure}

For any given relative rotation angle $\Delta\Theta$, we can numerically calculate the minimum and maximum values of $I'(\varphi)$,
and then compute the polarization degree with Eq. (\ref{eq:polarization2}). The observed polarization degree $\Pi'$ as a function of
$\Delta\Theta$ is displayed in Fig.~\ref{fig:f1}.
In Fig.~\ref{fig:f1}(a), the photon index is fixed at $\Gamma=-1.5$. Solid lines for different energy bands are shown: $100-300$,
$200-400$, $70-300$, and $50-500$ keV.
One can see that although the net polarization degree $\Pi'$ decreases rapidly as $\Delta\Theta$ increases at $\Delta\Theta\leq \pi$,
more than $60\%$ of the initial polarization degree can still be conserved at $\Delta\Theta=\pi/2$. The energy dependence of polarization begins to be evident
at $\Delta\Theta\simeq\pi$. Comparing the violet ($100-300$ keV) and red ($200-400$ keV) solid lines, we can see that
for the same energy width, high-energy photons have slightly smaller polarization than low-energy ones at a fixed
$\Delta\Theta$. This is because high-energy photons rotate a larger polarization angle $\Delta\theta$ than
low-energy ones propagating the same distance [see Eq.~(\ref{eq:delta-theta})], leading to the polarization direction
of high-energy photons being more mixed and their polarization degree being suppressed. Comparing the violet ($100-300$ keV),
green ($70-300$ keV), and blue ($50-500$ keV) solid lines, one may see that at a fixed $\Delta\Theta$, photons
in a wider energy band have larger polarization. However, it does not mean that larger polarization can indeed be detected
in a wider energy band. This is because a wider energy band will also lead to a larger rotation angle.

Next, we investigate the dependence of polarization degree on the photon index $\Gamma$ by fixing the energy band in
50--500 keV. The evolution of polarization for different values of $\Gamma$ are shown as solid lines in Fig.~\ref{fig:f1}(b).
We find that the net polarization degree increases with decreasing $\Gamma$ values. This is easy to understand. When $\Gamma$
is smaller, the photons are closer to monochromatic. In the strictly monochromatic case, the polarization degree has been proved to be
invariant during the propagation \citep{2016MNRAS.463..375L}.

\begin{figure}
\vskip-0.3in
\centerline{\includegraphics[angle=0,width=1.0\hsize]{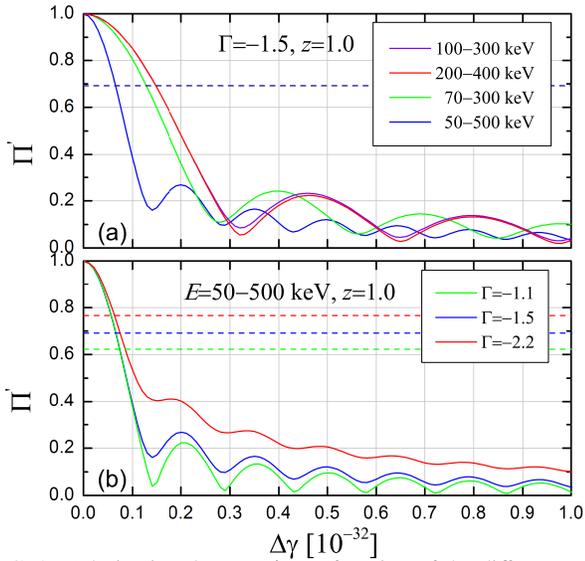}}
\vskip-0.4in
\caption{Polarization degree $\Pi'$ as a function of the difference of the $\gamma$ values (solid lines).
(a): The photon index $\Gamma$ is fixed at $-1.5$. (b): The energy band is fixed in 50--500 keV.
In both panels, the redshift of the source is assumed to be $z=1.0$.
Dashed lines represent the level of polarization that is produced by synchrotron emission.}
\label{fig:f2}
\end{figure}

Since the $\gamma$ discrepancy ($\Delta\gamma$) is related to $\Delta\Theta$, we can also
plot the polarization degree as a function of $\Delta\gamma$ in Fig.~\ref{fig:f2}. Note that the limits on $\Delta\gamma$ are not greatly affected by uncertainties
in the source distance and direction, as one can see from Eq.~(\ref{eq:gammadiff}).
Here the source is assumed to be located at a redshift $z=1.0$ (corresponding to a comoving distance of $d=3.4$ Gpc),
and the impact parameter $b$ (which $\leq r_L$) and the sign of the correction of the source direction $s_n$ are respectively
taken to be $b=0.5r_L$ and $s_n=+1$. In Fig.~\ref{fig:f2}(a), we present the evolution of polarization in different energy bands
but with the fixed photon index $\Gamma=-1.5$ (solid lines).
At the large $\Delta\gamma$ scale, where the polarization degree is below $\sim20\%$,
curves in different energy bands intersect with each other.
At small $\Delta\gamma$, the net polarization decreases rapidly with increasing $\Delta\gamma$.
A wider and higher energy band shows a steeper decline. Therefore, to tightly test the WEP accuracy (i.e., $\Delta\gamma$),
the polarization observation in a wider and higher energy band is favored.

In Fig.~\ref{fig:f2}(b), we show the polarization for various values of $\Gamma$
but with the fixed energy band $50-500$ keV (solid lines). Similar to the bottom panel of the $\Pi'-\Delta\Theta$ plot, this plot also shows
that the polarization degree decreases more rapidly with larger $\Gamma$.
Therefore, we can conclude that GRBs with harder spectra (i.e., larger $\Gamma$ values) are helpful to tightly constrain the WEP violation.

Several emission mechanisms (e.g., synchrotron emission) proposed for
GRB prompt emission may produce linear polarization degree as high as $\prod=(-\Gamma)/(-\Gamma+2/3)$,
independent of the energy band \citep{2012ApJ...758L...1Y}.
In Figs. \ref{fig:f1} and \ref{fig:f2}, we also plot the level of polarization
that synchrotron emission would be assumed to produce with different photon indices $\Gamma$ (dashed lines).
By comparing the solid and dashed lines, we can see that the polarization evolution arising from the WEP violation
is quite different from that is produced by synchrotron emission.

\section{WEP Test with GRB Polarization}

As mentioned above, the higher and wider the energy band of polarization observation and the harder the spectrum of the source,
the better the constraint is on the WEP violation. There have been some polarization measurements in the prompt gamma-ray emission of GRBs
(see \citet{2017NewAR..76....1M} for a recent review). We apply our formulas to two highly polarized bursts (GRB 110721A and GRB 061122)
and thereby place the strictest limits on a deviation from the WEP. These two bursts are selected as they have hard
spectra and show clear polarization signatures with high detection significance.

\begin{figure}
\vskip-0.2in
\centerline{\includegraphics[angle=0,width=1.1\hsize]{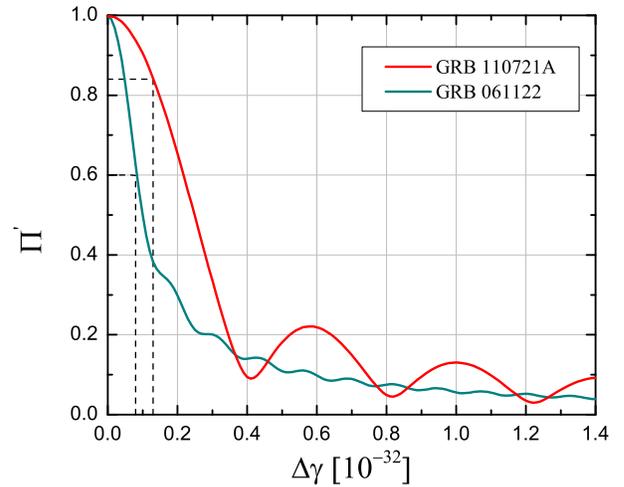}}
\vskip-0.1in
\caption{Limits on the differences of the $\gamma$ values from the polarization measurements of GRB 110721A and GRB 061122.
The vertical and horizonal dashed lines mark the $\Delta\gamma$ values that corresponds to the observed polarization degrees.}
\label{fig:f3}
\end{figure}

GRB 110721A was detected on July 21, 2011, with coordinates (J2000)
R.A.=$331.42^{\circ}$ and Dec.=$-36.42^{\circ}$ \citep{2011GCN.12187....1T}. Its redshift has been measured to be
$z=0.382$ \citep{2011GCN.12193....1B}. \citet{2012ApJ...758L...1Y} reported the detection of a linear polarization degree of
$\Pi=84^{+16}_{-28}\%$ in the energy range of the Gamma-Ray Burst Polarimeter ($70-300$ keV). The time-averaged spectrum can be well fitted by
a Band function with $E_p=372.50^{+26.50}_{-23.60}$ keV, $\alpha=-0.94\pm0.02$, and $\beta=-1.77\pm0.02$
\citep{2011GCN.12187....1T}. Using the above information, and considering the error propagation
from the measured polarization degree, we plot the evolution of polarization
degree in Fig.~\ref{fig:f3} and thereby obtain a conservative upper limit of the WEP violation, i.e.,
$\Delta\gamma<1.3^{+1.0}_{-1.3}\times10^{-33}$, which is $10^{6}$ times tighter than previous limits.

GRB 061122 was detected on 22 November 2006, with coordinates
R.A.=$20^{\rm h}15^{\rm m}20^{\rm s}.88$ and Dec.=$+15^{\circ}30^{'}50^{''}.8$ \citep{2006GCN..5834....1M}.
Its redshift is $z=1.33$. \citet{2013MNRAS.431.3550G}
measured linear polarization in the energy band ($250-800$ keV) during the prompt emission of
GRB 061122 and set a lower limit on its polarization degree of $\Pi>60\%$ at a $1\sigma$
confidence level. The spectrum can be well fitted by a power law with a high-energy cutoff,
i.e., $N(E)\propto E^{\alpha}\exp(-E/E_c)$, where $\alpha=-1.15\pm0.04$ and
$E_c=221\pm20$ keV. Using these observational values, the most conservative upper limit on the WEP violation is
$\Delta\gamma<0.8\times10^{-33}$ (see Fig.~\ref{fig:f3}),
which improves existing bounds on a deviation from the WEP by at least 6 orders of magnitude.
It should be underlined that these two constraints are obtained by assuming that the initial polarization degree is 100\%.
If photons are not initially completely polarized, the WEP test may be much tighter (see \citet{2016MNRAS.463..375L}
for detailed descriptions in the initially partially polarized case).

In the following, we discuss some possible systematic and astrophysical effects on our constraint results.

The largest systematic uncertainty comes from the determination of the gravitational potential of the Laniakea supercluster
of galaxies $U(r)$. The exact gravitational potential function of Laniakea is not well known. More accurate understanding of
the function $U(r)$ could improve our constraints, but this correction should be limited to less than an order of magnitude.

On the other hand, there are some astrophysical effects that affect the propagation of polarized gamma-ray emission of GRBs,
such as Faraday rotation, vacuum birefringence, and axion electrodynamics.

\begin{itemize}
  \item The rotation of the linear polarization angle would be
affected by magnetized plasmas (the so-called Faraday rotation). The dependence of the rotation angle on
Faraday rotation is $\Delta\theta_{\rm Far}\propto E^{-2}$, different from its dependence on the WEP violation as
$\Delta\theta_{\rm WEP}\propto E$ shown in Eq.(\ref{eq:delta-theta}).
For high-energy photons, such as the gamma-ray signals considered here, $\Delta\theta_{\rm Far}$ is negligible
for the purposes of this work.
  \item  Many quantum gravity theories predict that Lorentz invariance may be broken.
Lorentz invariance violation (LIV) can produce vacuum birefringence, which leads to a phase rotation
of linear polarization. The rotation angle induced by the LIV effect is
$\Delta\theta_{\rm LIV}=\frac{\xi E^2}{M_{\rm pl}}\int_0^z\frac{(1+z')dz'}{H(z')}$, where $M_{\rm pl}$ is the Planck
energy and $H(z)$ is the Hubble parameter. \citet{2012PhRvL.109x1104T} has provided an upper limit for
the birefringence parameter as $\xi < \mathcal{O}(10^{-15})$. In this case, $\Delta\theta_{\rm LIV}$ is negligible
in our analysis.
  \item Axion dark matter also causes rotation of the polarization of light passing through it.
The rotation angle of the polarization plane arising from the axion--photon coupling ($\Delta\theta_{\rm axion}$)
is proportional to the difference of the axion field values at photon emission and photon absorption ($\Delta\phi$),
i.e., $\Delta\theta_{\rm axion}\propto \Delta\phi$ \citep{1992PhLB..289...67H}.
Applying this understanding to the polarized fraction of the CMB, \citet{2019arXiv190302666F}
found that there are two novel phenomena in the CMB, which can be used to search for
low-mass axion dark matter. The rotation effect of axion electrodynamics depends only on
the initial and final axion field values \citep{1992PhLB..289...67H}, which is quite different from the energy-dependent rotation one
induced by the WEP violation.
\end{itemize}

\section{Conclusions}
\label{sec:summary}
In this work, we have investigated the evolution of GRB polarization induced by a possible violation of the WEP.
Once the WEP fails, then different values of the PPN parameter $\gamma$ are measured with different circularly polarized
photons, which results in slightly different arrival times for different circular polarization states,
leading to an energy-dependent rotation of the polarization plane of a linear polarized wave.
We derive the net polarization degree as a function of the relative rotation angle $\Delta\Theta$,
where $\Delta\Theta$ denotes the difference of rotation angles over an energy range.
In contrast to the intuition that the net polarization will be severely suppressed when $\Delta\Theta$ is large,
we find that, even if $\Delta\Theta$ is as large as $\pi/2$, more than $60\%$ of the initial polarization degree can still be conserved.
Thus, it is inappropriate to simply use a fixed value as the upper limit of $\Delta\Theta$ to test the WEP accuracy.
In addition, we prove that GRBs with harder spectra and polarization observations in a wider energy range are more favorable to constrain the WEP violation.

Applying our formulas to the GRB polarimetric data and attributing the Shapiro delay to
the gravitational potential of the Laniakea supercluster of galaxies, we place the most stringent limits so far on $\gamma$ differences
for two cases: $\Delta\gamma<1.3^{+1.0}_{-1.3}\times10^{-33}$ for GRB 110721A and $\Delta\gamma<0.8\times10^{-33}$ for GRB 061122.
Previously, with the assumptions that the relative rotation angle $\Delta\Theta$ cannot exceed more than $2\pi$
and the Shapiro delay is caused by the Milky Way's gravity, \citet{2017MNRAS.469L..36Y} set a severe limit
of $\Delta\gamma<1.6\times10^{-27}$ from the polarization observation of GRB 110721A.
Compared with this previous best result, our limits represent an improvement of 6 orders of magnitude.
Although the WEP test can indeed be tightened by incorporating the Laniakea's gravity rather than the Milky Way's gravity,
we find that even if the Shapiro delay is only due to the Milky Way's gravity, a strict limit of
$\Delta\gamma<0.7^{+0.5}_{-0.7}\times10^{-28}$ from GRB 110721A can still be achieved using our method, which is already 10 times tighter than that obtained
by setting the upper limit of $\Delta\Theta$ to be $2\pi$ \citep{2017MNRAS.469L..36Y}.
If the GRB polarimetric data can be significantly enlarged in the future, a much more statistically robust bound on
a deviation from the WEP can be expected.

\begin{acknowledgments}
We are grateful to the anonymous referees for insightful comments.
This work is partially supported by the National Natural Science Foundation of China
(grant Nos. U1831122, 11603076, 11673068, and 11725314), the Youth Innovation Promotion
Association (2017366), the Key Research Program of Frontier Sciences (grant No. QYZDB-SSW-SYS005),
the Strategic Priority Research Program ``Multi-waveband gravitational wave Universe''
(grant No. XDB23000000) of Chinese Academy of Sciences, and the ``333 Project''
and the Natural Science Foundation (grant No. BK20161096) of Jiangsu Province.
\end{acknowledgments}


\begin{thebibliography}{48}
\expandafter\ifx\csname natexlab\endcsname\relax\def\natexlab#1{#1}\fi
\expandafter\ifx\csname bibnamefont\endcsname\relax
  \def\bibnamefont#1{#1}\fi
\expandafter\ifx\csname bibfnamefont\endcsname\relax
  \def\bibfnamefont#1{#1}\fi
\expandafter\ifx\csname citenamefont\endcsname\relax
  \def\citenamefont#1{#1}\fi
\expandafter\ifx\csname url\endcsname\relax
  \def\url#1{\texttt{#1}}\fi
\expandafter\ifx\csname urlprefix\endcsname\relax\def\urlprefix{URL }\fi
\providecommand{\bibinfo}[2]{#2}
\providecommand{\eprint}[2][]{\url{#2}}

\bibitem[{\citenamefont{{Will}}(2006)}]{2006LRR.....9....3W}
\bibinfo{author}{\bibfnamefont{C.~M.} \bibnamefont{{Will}}},
  \bibinfo{journal}{Living Rev. Rel.} \textbf{\bibinfo{volume}{9}},
  \bibinfo{eid}{3} (\bibinfo{year}{2006}); \bibinfo{author}{\bibfnamefont{C.~M.} \bibnamefont{{Will}}},
  \bibinfo{journal}{Living Rev. Rel.} \textbf{\bibinfo{volume}{17}},
  \bibinfo{eid}{4} (\bibinfo{year}{2014}).

\bibitem[{\citenamefont{{Shapiro}}(1964)}]{1964PhRvL..13..789S}
\bibinfo{author}{\bibfnamefont{I.~I.} \bibnamefont{{Shapiro}}},
  \bibinfo{journal}{Phys. Rev. Lett.} \textbf{\bibinfo{volume}{13}},
  \bibinfo{pages}{789} (\bibinfo{year}{1964}).

\bibitem[{\citenamefont{{Longo}}(1988)}]{1988PhRvL..60..173L}
\bibinfo{author}{\bibfnamefont{M.~J.} \bibnamefont{{Longo}}},
  \bibinfo{journal}{Phys. Rev. Lett.} \textbf{\bibinfo{volume}{60}},
  \bibinfo{pages}{173} (\bibinfo{year}{1988});
  \bibinfo{author}{\bibfnamefont{L.~M.} \bibnamefont{{Krauss}}} \bibnamefont{and}
  \bibinfo{author}{\bibfnamefont{S.}~\bibnamefont{{Tremaine}}},
  \bibinfo{journal}{Phys. Rev. Lett.} \textbf{\bibinfo{volume}{60}},
  \bibinfo{pages}{176} (\bibinfo{year}{1988});
  \bibinfo{author}{\bibfnamefont{X.-F.} \bibnamefont{{Wu}}},
  \bibinfo{author}{\bibfnamefont{H.}~\bibnamefont{{Gao}}},
  \bibinfo{author}{\bibfnamefont{J.-J.} \bibnamefont{{Wei}}},
  \bibinfo{author}{\bibfnamefont{P.}~\bibnamefont{{M{\'e}sz{\'a}ros}}},
  \bibinfo{author}{\bibfnamefont{B.}~\bibnamefont{{Zhang}}},
  \bibinfo{author}{\bibfnamefont{Z.-G.} \bibnamefont{{Dai}}},
  \bibinfo{author}{\bibfnamefont{S.-N.} \bibnamefont{{Zhang}}},
  \bibnamefont{and} \bibinfo{author}{\bibfnamefont{Z.-H.} \bibnamefont{{Zhu}}},
  \bibinfo{journal}{Phys. Rev. D} \textbf{\bibinfo{volume}{94}},
  \bibinfo{eid}{024061} (\bibinfo{year}{2016}).

\bibitem[{\citenamefont{{Wei} et~al.}(2016{\natexlab{a}})\citenamefont{{Wei},
  {Wu}, {Gao}, and {M{\'e}sz{\'a}ros}}}]{2016JCAP...08..031W}
\bibinfo{author}{\bibfnamefont{J.-J.} \bibnamefont{{Wei}}},
  \bibinfo{author}{\bibfnamefont{X.-F.} \bibnamefont{{Wu}}},
  \bibinfo{author}{\bibfnamefont{H.}~\bibnamefont{{Gao}}}, \bibnamefont{and}
  \bibinfo{author}{\bibfnamefont{P.}~\bibnamefont{{M{\'e}sz{\'a}ros}}},
  \bibinfo{journal}{JCAP} \textbf{\bibinfo{volume}{8}}, \bibinfo{eid}{031}
  (\bibinfo{year}{2016}{\natexlab{a}});
  \bibinfo{author}{\bibfnamefont{H.}~\bibnamefont{{Yu}}},
  \bibinfo{author}{\bibfnamefont{S.-Q.} \bibnamefont{{Xi}}}, \bibnamefont{and}
  \bibinfo{author}{\bibfnamefont{F.-Y.} \bibnamefont{{Wang}}},
  \bibinfo{journal}{Astrophys. J.} \textbf{\bibinfo{volume}{860}},
  \bibinfo{eid}{173} (\bibinfo{year}{2018});
      \bibinfo{author}{\bibfnamefont{R.}~\bibnamefont{{Laha}}},
  \bibinfo{journal}{ArXiv e-prints}  (\bibinfo{year}{2018}),
  \eprint{1807.05621};
  \bibinfo{author}{\bibfnamefont{J.-J.} \bibnamefont{{Wei}}},
  \bibinfo{author}{\bibfnamefont{B.-B.} \bibnamefont{{Zhang}}},
  \bibinfo{author}{\bibfnamefont{L.}~\bibnamefont{{Shao}}},
  \bibinfo{author}{\bibfnamefont{H.}~\bibnamefont{{Gao}}},
  \bibinfo{author}{\bibfnamefont{Y.}~\bibnamefont{{Li}}},
  \bibinfo{author}{\bibfnamefont{Q.-Q.} \bibnamefont{{Yin}}},
  \bibinfo{author}{\bibfnamefont{X.-F.} \bibnamefont{{Wu}}},
  \bibinfo{author}{\bibfnamefont{X.-Y.} \bibnamefont{{Wang}}},
  \bibinfo{author}{\bibfnamefont{B.}~\bibnamefont{{Zhang}}}, \bibnamefont{and}
  \bibinfo{author}{\bibfnamefont{Z.-G.} \bibnamefont{{Dai}}},
  \bibinfo{journal}{Journal of High Energy Astrophysics}
  \textbf{\bibinfo{volume}{22}}, \bibinfo{pages}{1} (\bibinfo{year}{2019}).

\bibitem[{\citenamefont{{Gao} et~al.}(2015)\citenamefont{{Gao}, {Wu}, and
  {M{\'e}sz{\'a}ros}}}]{2015ApJ...810..121G}
\bibinfo{author}{\bibfnamefont{H.}~\bibnamefont{{Gao}}},
  \bibinfo{author}{\bibfnamefont{X.-F.} \bibnamefont{{Wu}}}, \bibnamefont{and}
  \bibinfo{author}{\bibfnamefont{P.}~\bibnamefont{{M{\'e}sz{\'a}ros}}},
  \bibinfo{journal}{Astrophys. J.} \textbf{\bibinfo{volume}{810}},
  \bibinfo{eid}{121} (\bibinfo{year}{2015});
  \bibinfo{author}{\bibfnamefont{J.-J.} \bibnamefont{{Wei}}},
  \bibinfo{author}{\bibfnamefont{H.}~\bibnamefont{{Gao}}},
  \bibinfo{author}{\bibfnamefont{X.-F.} \bibnamefont{{Wu}}}, \bibnamefont{and}
  \bibinfo{author}{\bibfnamefont{P.}~\bibnamefont{{M{\'e}sz{\'a}ros}}},
  \bibinfo{journal}{Phys. Rev. Lett.} \textbf{\bibinfo{volume}{115}},
  \bibinfo{eid}{261101} (\bibinfo{year}{2015});
  \bibinfo{author}{\bibfnamefont{J.-J.} \bibnamefont{{Wei}}},
  \bibinfo{author}{\bibfnamefont{J.-S.} \bibnamefont{{Wang}}},
  \bibinfo{author}{\bibfnamefont{H.}~\bibnamefont{{Gao}}}, \bibnamefont{and}
  \bibinfo{author}{\bibfnamefont{X.-F.} \bibnamefont{{Wu}}},
  \bibinfo{journal}{Astrophys. J.} \textbf{\bibinfo{volume}{818}},
  \bibinfo{eid}{L2} (\bibinfo{year}{2016}{\natexlab{b}});
  \bibinfo{author}{\bibfnamefont{Y.}~\bibnamefont{{Sang}}},
  \bibinfo{author}{\bibfnamefont{H.-N.} \bibnamefont{{Lin}}}, \bibnamefont{and}
  \bibinfo{author}{\bibfnamefont{Z.}~\bibnamefont{{Chang}}},
  \bibinfo{journal}{Mon. Not. R. Astron. Soc.} \textbf{\bibinfo{volume}{460}},
  \bibinfo{pages}{2282} (\bibinfo{year}{2016});
  \bibinfo{author}{\bibfnamefont{S.~J.} \bibnamefont{{Tingay}}} \bibnamefont{and}
  \bibinfo{author}{\bibfnamefont{D.~L.} \bibnamefont{{Kaplan}}},
  \bibinfo{journal}{Astrophys. J.} \textbf{\bibinfo{volume}{820}},
  \bibinfo{eid}{L31} (\bibinfo{year}{2016});
  \bibinfo{author}{\bibfnamefont{Z.-Y.} \bibnamefont{{Wang}}},
  \bibinfo{author}{\bibfnamefont{R.-Y.} \bibnamefont{{Liu}}}, \bibnamefont{and}
  \bibinfo{author}{\bibfnamefont{X.-Y.} \bibnamefont{{Wang}}},
  \bibinfo{journal}{Phys. Rev. Lett.} \textbf{\bibinfo{volume}{116}},
  \bibinfo{eid}{151101} (\bibinfo{year}{2016});
\bibinfo{author}{\bibfnamefont{Y.-P.} \bibnamefont{{Yang}}} \bibnamefont{and}
  \bibinfo{author}{\bibfnamefont{B.}~\bibnamefont{{Zhang}}},
  \bibinfo{journal}{Phys. Rev. D} \textbf{\bibinfo{volume}{94}},
  \bibinfo{eid}{101501} (\bibinfo{year}{2016});
  \bibinfo{author}{\bibfnamefont{Y.}~\bibnamefont{{Zhang}}} \bibnamefont{and}
  \bibinfo{author}{\bibfnamefont{B.}~\bibnamefont{{Gong}}},
  \bibinfo{journal}{Astrophys. J.} \textbf{\bibinfo{volume}{837}},
  \bibinfo{eid}{134} (\bibinfo{year}{2017});
  \bibinfo{author}{\bibfnamefont{S.}~\bibnamefont{{Desai}}} \bibnamefont{and}
  \bibinfo{author}{\bibfnamefont{E.}~\bibnamefont{{Kahya}}},
  \bibinfo{journal}{European Physical Journal C} \textbf{\bibinfo{volume}{78}},
  \bibinfo{eid}{86} (\bibinfo{year}{2018});
  \bibinfo{author}{\bibfnamefont{C.}~\bibnamefont{{Leung}}},
  \bibinfo{author}{\bibfnamefont{B.}~\bibnamefont{{Hu}}},
  \bibinfo{author}{\bibfnamefont{S.}~\bibnamefont{{Harris}}},
  \bibinfo{author}{\bibfnamefont{A.}~\bibnamefont{{Brown}}},
  \bibinfo{author}{\bibfnamefont{J.}~\bibnamefont{{Gallicchio}}},
  \bibnamefont{and} \bibinfo{author}{\bibfnamefont{H.}~\bibnamefont{{Nguyen}}},
  \bibinfo{journal}{Astrophys. J.} \textbf{\bibinfo{volume}{861}},
  \bibinfo{eid}{66} (\bibinfo{year}{2018});
\bibinfo{author}{\bibfnamefont{E.~O.} \bibnamefont{{Kahya}}} \bibnamefont{and}
  \bibinfo{author}{\bibfnamefont{S.}~\bibnamefont{{Desai}}},
  \bibinfo{journal}{Phys. Lett. B} \textbf{\bibinfo{volume}{756}},
  \bibinfo{pages}{265} (\bibinfo{year}{2016});
  \bibinfo{author}{\bibfnamefont{X.}~\bibnamefont{{Li}}},
  \bibinfo{author}{\bibfnamefont{Y.-M.} \bibnamefont{{Hu}}},
  \bibinfo{author}{\bibfnamefont{Y.-Z.} \bibnamefont{{Fan}}}, \bibnamefont{and}
  \bibinfo{author}{\bibfnamefont{D.-M.} \bibnamefont{{Wei}}},
  \bibinfo{journal}{Astrophys. J.} \textbf{\bibinfo{volume}{827}},
  \bibinfo{eid}{75} (\bibinfo{year}{2016});
  \bibinfo{author}{\bibfnamefont{M.}~\bibnamefont{{Liu}}},
  \bibinfo{author}{\bibfnamefont{Z.}~\bibnamefont{{Zhao}}},
  \bibinfo{author}{\bibfnamefont{X.}~\bibnamefont{{You}}},
  \bibinfo{author}{\bibfnamefont{J.}~\bibnamefont{{Lu}}}, \bibnamefont{and}
  \bibinfo{author}{\bibfnamefont{L.}~\bibnamefont{{Xu}}},
  \bibinfo{journal}{Phys. Lett. B} \textbf{\bibinfo{volume}{770}},
  \bibinfo{pages}{8} (\bibinfo{year}{2017});
  \bibinfo{author}{\bibfnamefont{J.-J.} \bibnamefont{{Wei}}},
  \bibinfo{author}{\bibfnamefont{B.-B.} \bibnamefont{{Zhang}}},
  \bibinfo{author}{\bibfnamefont{X.-F.} \bibnamefont{{Wu}}},
  \bibinfo{author}{\bibfnamefont{H.}~\bibnamefont{{Gao}}},
  \bibinfo{author}{\bibfnamefont{P.}~\bibnamefont{{M{\'e}sz{\'a}ros}}},
  \bibinfo{author}{\bibfnamefont{B.}~\bibnamefont{{Zhang}}},
  \bibinfo{author}{\bibfnamefont{Z.-G.} \bibnamefont{{Dai}}},
  \bibinfo{author}{\bibfnamefont{S.-N.} \bibnamefont{{Zhang}}},
  \bibnamefont{and} \bibinfo{author}{\bibfnamefont{Z.-H.} \bibnamefont{{Zhu}}},
  \bibinfo{journal}{JCAP} \textbf{\bibinfo{volume}{11}}, \bibinfo{eid}{035}
  (\bibinfo{year}{2017});
  \bibinfo{author}{\bibfnamefont{B.~P.} \bibnamefont{{Abbott}}},
  \bibinfo{author}{\bibfnamefont{R.}~\bibnamefont{{Abbott}}},
  \bibinfo{author}{\bibfnamefont{T.~D.} \bibnamefont{{Abbott}}},
  \bibinfo{author}{\bibfnamefont{F.}~\bibnamefont{{Acernese}}},
  \bibinfo{author}{\bibfnamefont{K.}~\bibnamefont{{Ackley}}},
  \bibinfo{author}{\bibfnamefont{C.}~\bibnamefont{{Adams}}},
  \bibinfo{author}{\bibfnamefont{T.}~\bibnamefont{{Adams}}},
  \bibinfo{author}{\bibfnamefont{P.}~\bibnamefont{{Addesso}}},
  \bibinfo{author}{\bibfnamefont{R.~X.} \bibnamefont{{Adhikari}}},
  \bibinfo{author}{\bibfnamefont{V.~B.} \bibnamefont{{Adya}}},
  \bibnamefont{et~al.}, \bibinfo{journal}{Astrophys. J.}
  \textbf{\bibinfo{volume}{848}}, \bibinfo{eid}{L13} (\bibinfo{year}{2017});
  \bibinfo{author}{\bibfnamefont{I.~M.} \bibnamefont{{Shoemaker}}}
  \bibnamefont{and} \bibinfo{author}{\bibfnamefont{K.}~\bibnamefont{{Murase}}},
  \bibinfo{journal}{\prd} \textbf{\bibinfo{volume}{97}}, \bibinfo{eid}{083013}
  (\bibinfo{year}{2018});
    \bibinfo{author}{\bibfnamefont{S.}~\bibnamefont{{Boran}}},
  \bibinfo{author}{\bibfnamefont{S.}~\bibnamefont{{Desai}}}, \bibnamefont{and}
  \bibinfo{author}{\bibfnamefont{E.~O.} \bibnamefont{{Kahya}}},
    \bibinfo{journal}{European Physical Journal C} \textbf{\bibinfo{volume}{79}},
  \bibinfo{eid}{185} (\bibinfo{year}{2019}).


\bibitem[{\citenamefont{{Wu} et~al.}(2017)\citenamefont{{Wu}, {Wei}, {Lan},
  {Gao}, {Dai}, and {M{\'e}sz{\'a}ros}}}]{2017PhRvD..95j3004W}
\bibinfo{author}{\bibfnamefont{X.-F.} \bibnamefont{{Wu}}},
  \bibinfo{author}{\bibfnamefont{J.-J.} \bibnamefont{{Wei}}},
  \bibinfo{author}{\bibfnamefont{M.-X.} \bibnamefont{{Lan}}},
  \bibinfo{author}{\bibfnamefont{H.}~\bibnamefont{{Gao}}},
  \bibinfo{author}{\bibfnamefont{Z.-G.} \bibnamefont{{Dai}}}, \bibnamefont{and}
  \bibinfo{author}{\bibfnamefont{P.}~\bibnamefont{{M{\'e}sz{\'a}ros}}},
  \bibinfo{journal}{Phys. Rev. D} \textbf{\bibinfo{volume}{95}},
  \bibinfo{eid}{103004} (\bibinfo{year}{2017}).

\bibitem[{\citenamefont{{Toma} et~al.}(2012)\citenamefont{{Toma}, {Mukohyama},
  {Yonetoku}, {Murakami}, {Gunji}, {Mihara}, {Morihara}, {Sakashita},
  {Takahashi}, {Wakashima} et~al.}}]{2012PhRvL.109x1104T}
\bibinfo{author}{\bibfnamefont{K.}~\bibnamefont{{Toma}}},
  \bibinfo{author}{\bibfnamefont{S.}~\bibnamefont{{Mukohyama}}},
  \bibinfo{author}{\bibfnamefont{D.}~\bibnamefont{{Yonetoku}}},
  \bibinfo{author}{\bibfnamefont{T.}~\bibnamefont{{Murakami}}},
  \bibinfo{author}{\bibfnamefont{S.}~\bibnamefont{{Gunji}}},
  \bibinfo{author}{\bibfnamefont{T.}~\bibnamefont{{Mihara}}},
  \bibinfo{author}{\bibfnamefont{Y.}~\bibnamefont{{Morihara}}},
  \bibinfo{author}{\bibfnamefont{T.}~\bibnamefont{{Sakashita}}},
  \bibinfo{author}{\bibfnamefont{T.}~\bibnamefont{{Takahashi}}},
  \bibinfo{author}{\bibfnamefont{Y.}~\bibnamefont{{Wakashima}}},
  \bibnamefont{et~al.}, \bibinfo{journal}{Phys. Rev. Lett.}
  \textbf{\bibinfo{volume}{109}}, \bibinfo{eid}{241104} (\bibinfo{year}{2012}).

\bibitem[{\citenamefont{{Yang} et~al.}(2017)\citenamefont{{Yang}, {Zou},
  {Zhang}, {Liao}, and {Lei}}}]{2017MNRAS.469L..36Y}
\bibinfo{author}{\bibfnamefont{C.}~\bibnamefont{{Yang}}},
  \bibinfo{author}{\bibfnamefont{Y.-C.} \bibnamefont{{Zou}}},
  \bibinfo{author}{\bibfnamefont{Y.-Y.} \bibnamefont{{Zhang}}},
  \bibinfo{author}{\bibfnamefont{B.}~\bibnamefont{{Liao}}}, \bibnamefont{and}
  \bibinfo{author}{\bibfnamefont{W.-H.} \bibnamefont{{Lei}}},
  \bibinfo{journal}{Mon. Not. R. Astron. Soc.} \textbf{\bibinfo{volume}{469}},
  \bibinfo{pages}{L36} (\bibinfo{year}{2017}).

\bibitem[{\citenamefont{{Luo} et~al.}(2016)\citenamefont{{Luo}, {Zhang}, {Wei},
  and {Wu}}}]{2016JHEAp...9...35L}
\bibinfo{author}{\bibfnamefont{Z.-X.} \bibnamefont{{Luo}}},
  \bibinfo{author}{\bibfnamefont{B.}~\bibnamefont{{Zhang}}},
  \bibinfo{author}{\bibfnamefont{J.-J.} \bibnamefont{{Wei}}}, \bibnamefont{and}
  \bibinfo{author}{\bibfnamefont{X.-F.} \bibnamefont{{Wu}}},
  \bibinfo{journal}{JHEAp} \textbf{\bibinfo{volume}{9}}, \bibinfo{pages}{35}
  (\bibinfo{year}{2016});
  \bibinfo{author}{\bibfnamefont{A.}~\bibnamefont{{Nusser}}},
  \bibinfo{journal}{Astrophys. J.} \textbf{\bibinfo{volume}{821}},
  \bibinfo{eid}{L2} (\bibinfo{year}{2016});
  \bibinfo{author}{\bibfnamefont{S.-N.} \bibnamefont{{Zhang}}},
  \bibinfo{journal}{ArXiv e-prints}  (\bibinfo{year}{2016}),
  \eprint{1601.04558}.

\bibitem[{\citenamefont{{Tully} et~al.}(2014)\citenamefont{{Tully}, {Courtois},
  {Hoffman}, and {Pomar{\`e}de}}}]{2014Natur.513...71T}
\bibinfo{author}{\bibfnamefont{R.~B.} \bibnamefont{{Tully}}},
  \bibinfo{author}{\bibfnamefont{H.}~\bibnamefont{{Courtois}}},
  \bibinfo{author}{\bibfnamefont{Y.}~\bibnamefont{{Hoffman}}},
  \bibnamefont{and}
  \bibinfo{author}{\bibfnamefont{D.}~\bibnamefont{{Pomar{\`e}de}}},
  \bibinfo{journal}{Nature} \textbf{\bibinfo{volume}{513}}, \bibinfo{pages}{71}
  (\bibinfo{year}{2014}).

\bibitem[{\citenamefont{{Lynden-Bell} et~al.}(1988)\citenamefont{{Lynden-Bell},
  {Faber}, {Burstein}, {Davies}, {Dressler}, {Terlevich}, and
  {Wegner}}}]{1988ApJ...326...19L}
\bibinfo{author}{\bibfnamefont{D.}~\bibnamefont{{Lynden-Bell}}},
  \bibinfo{author}{\bibfnamefont{S.~M.} \bibnamefont{{Faber}}},
  \bibinfo{author}{\bibfnamefont{D.}~\bibnamefont{{Burstein}}},
  \bibinfo{author}{\bibfnamefont{R.~L.} \bibnamefont{{Davies}}},
  \bibinfo{author}{\bibfnamefont{A.}~\bibnamefont{{Dressler}}},
  \bibinfo{author}{\bibfnamefont{R.~J.} \bibnamefont{{Terlevich}}},
  \bibnamefont{and} \bibinfo{author}{\bibfnamefont{G.}~\bibnamefont{{Wegner}}},
  \bibinfo{journal}{Astrophys. J.} \textbf{\bibinfo{volume}{326}},
  \bibinfo{pages}{19} (\bibinfo{year}{1988}).

\bibitem[{\citenamefont{{Lin} et~al.}(2016)\citenamefont{{Lin}, {Li}, and
  {Chang}}}]{2016MNRAS.463..375L}
\bibinfo{author}{\bibfnamefont{H.-N.} \bibnamefont{{Lin}}},
  \bibinfo{author}{\bibfnamefont{X.}~\bibnamefont{{Li}}}, \bibnamefont{and}
  \bibinfo{author}{\bibfnamefont{Z.}~\bibnamefont{{Chang}}},
  \bibinfo{journal}{\mnras} \textbf{\bibinfo{volume}{463}},
  \bibinfo{pages}{375} (\bibinfo{year}{2016}).

\bibitem[{\citenamefont{{Rybicki} and {Lightman}}(1979)}]{1979rpa..book.....R}
\bibinfo{author}{\bibfnamefont{G.~B.} \bibnamefont{{Rybicki}}}
  \bibnamefont{and} \bibinfo{author}{\bibfnamefont{A.~P.}
  \bibnamefont{{Lightman}}}, \emph{\bibinfo{title}{{Radiative processes in
  astrophysics}}} (\bibinfo{year}{1979}).

\bibitem[{\citenamefont{{Band} et~al.}(1993)\citenamefont{{Band}, {Matteson},
  {Ford}, {Schaefer}, {Palmer}, {Teegarden}, {Cline}, {Briggs}, {Paciesas},
  {Pendleton} et~al.}}]{1993ApJ...413..281B}
\bibinfo{author}{\bibfnamefont{D.}~\bibnamefont{{Band}}},
  \bibinfo{author}{\bibfnamefont{J.}~\bibnamefont{{Matteson}}},
  \bibinfo{author}{\bibfnamefont{L.}~\bibnamefont{{Ford}}},
  \bibinfo{author}{\bibfnamefont{B.}~\bibnamefont{{Schaefer}}},
  \bibinfo{author}{\bibfnamefont{D.}~\bibnamefont{{Palmer}}},
  \bibinfo{author}{\bibfnamefont{B.}~\bibnamefont{{Teegarden}}},
  \bibinfo{author}{\bibfnamefont{T.}~\bibnamefont{{Cline}}},
  \bibinfo{author}{\bibfnamefont{M.}~\bibnamefont{{Briggs}}},
  \bibinfo{author}{\bibfnamefont{W.}~\bibnamefont{{Paciesas}}},
  \bibinfo{author}{\bibfnamefont{G.}~\bibnamefont{{Pendleton}}},
  \bibnamefont{et~al.}, \bibinfo{journal}{\apj} \textbf{\bibinfo{volume}{413}},
  \bibinfo{pages}{281} (\bibinfo{year}{1993}).

\bibitem[{\citenamefont{{Preece} et~al.}(2000)\citenamefont{{Preece}, {Briggs},
  {Mallozzi}, {Pendleton}, {Paciesas}, and {Band}}}]{2000ApJS..126...19P}
\bibinfo{author}{\bibfnamefont{R.~D.} \bibnamefont{{Preece}}},
  \bibinfo{author}{\bibfnamefont{M.~S.} \bibnamefont{{Briggs}}},
  \bibinfo{author}{\bibfnamefont{R.~S.} \bibnamefont{{Mallozzi}}},
  \bibinfo{author}{\bibfnamefont{G.~N.} \bibnamefont{{Pendleton}}},
  \bibinfo{author}{\bibfnamefont{W.~S.} \bibnamefont{{Paciesas}}},
  \bibnamefont{and} \bibinfo{author}{\bibfnamefont{D.~L.}
  \bibnamefont{{Band}}}, \bibinfo{journal}{\apjs}
  \textbf{\bibinfo{volume}{126}}, \bibinfo{pages}{19} (\bibinfo{year}{2000}).

\bibitem[{\citenamefont{{Yonetoku} et~al.}(2012)\citenamefont{{Yonetoku},
  {Murakami}, {Gunji}, {Mihara}, {Toma}, {Morihara}, {Takahashi}, {Wakashima},
  {Yonemochi}, {Sakashita} et~al.}}]{2012ApJ...758L...1Y}
\bibinfo{author}{\bibfnamefont{D.}~\bibnamefont{{Yonetoku}}},
  \bibinfo{author}{\bibfnamefont{T.}~\bibnamefont{{Murakami}}},
  \bibinfo{author}{\bibfnamefont{S.}~\bibnamefont{{Gunji}}},
  \bibinfo{author}{\bibfnamefont{T.}~\bibnamefont{{Mihara}}},
  \bibinfo{author}{\bibfnamefont{K.}~\bibnamefont{{Toma}}},
  \bibinfo{author}{\bibfnamefont{Y.}~\bibnamefont{{Morihara}}},
  \bibinfo{author}{\bibfnamefont{T.}~\bibnamefont{{Takahashi}}},
  \bibinfo{author}{\bibfnamefont{Y.}~\bibnamefont{{Wakashima}}},
  \bibinfo{author}{\bibfnamefont{H.}~\bibnamefont{{Yonemochi}}},
  \bibinfo{author}{\bibfnamefont{T.}~\bibnamefont{{Sakashita}}},
  \bibnamefont{et~al.}, \bibinfo{journal}{\apjl}
  \textbf{\bibinfo{volume}{758}}, \bibinfo{eid}{L1} (\bibinfo{year}{2012}).

\bibitem[{\citenamefont{{McConnell}}(2017)}]{2017NewAR..76....1M}
\bibinfo{author}{\bibfnamefont{M.~L.} \bibnamefont{{McConnell}}},
  \bibinfo{journal}{New A. Rev.} \textbf{\bibinfo{volume}{76}},
  \bibinfo{pages}{1} (\bibinfo{year}{2017}).

\bibitem[{\citenamefont{{Tierney} and {von
  Kienlin}}(2011)}]{2011GCN.12187....1T}
\bibinfo{author}{\bibfnamefont{D.}~\bibnamefont{{Tierney}}} \bibnamefont{and}
  \bibinfo{author}{\bibfnamefont{A.}~\bibnamefont{{von Kienlin}}},
  \bibinfo{journal}{GRB Coordinates Network, Circular Service, No.~12187, \#1
  (2011)} \textbf{\bibinfo{volume}{12187}} (\bibinfo{year}{2011}).

\bibitem[{\citenamefont{{Berger}}(2011)}]{2011GCN.12193....1B}
\bibinfo{author}{\bibfnamefont{E.}~\bibnamefont{{Berger}}},
  \bibinfo{journal}{GRB Coordinates Network, Circular Service, No.~12193, \#1
  (2011)} \textbf{\bibinfo{volume}{12193}} (\bibinfo{year}{2011}).


\bibitem[{\citenamefont{{Mereghetti} et~al.}(2006)\citenamefont{{Mereghetti},
  {Paizis}, {Gotz}, {Petry}, {Mowlavi}, {Beck}, and
  {Borkowski}}}]{2006GCN..5834....1M}
\bibinfo{author}{\bibfnamefont{S.}~\bibnamefont{{Mereghetti}}},
  \bibinfo{author}{\bibfnamefont{A.}~\bibnamefont{{Paizis}}},
  \bibinfo{author}{\bibfnamefont{D.}~\bibnamefont{{Gotz}}},
  \bibinfo{author}{\bibfnamefont{D.}~\bibnamefont{{Petry}}},
  \bibinfo{author}{\bibfnamefont{N.}~\bibnamefont{{Mowlavi}}},
  \bibinfo{author}{\bibfnamefont{M.}~\bibnamefont{{Beck}}}, \bibnamefont{and}
  \bibinfo{author}{\bibfnamefont{J.}~\bibnamefont{{Borkowski}}},
  \bibinfo{journal}{GRB Coordinates Network} \textbf{\bibinfo{volume}{5834}}
  (\bibinfo{year}{2006}).

\bibitem[{\citenamefont{{G{\"o}tz} et~al.}(2013)\citenamefont{{G{\"o}tz},
  {Covino}, {Fern{\'a}ndez-Soto}, {Laurent}, and {Bo{\v
  s}njak}}}]{2013MNRAS.431.3550G}
\bibinfo{author}{\bibfnamefont{D.}~\bibnamefont{{G{\"o}tz}}},
  \bibinfo{author}{\bibfnamefont{S.}~\bibnamefont{{Covino}}},
  \bibinfo{author}{\bibfnamefont{A.}~\bibnamefont{{Fern{\'a}ndez-Soto}}},
  \bibinfo{author}{\bibfnamefont{P.}~\bibnamefont{{Laurent}}},
  \bibnamefont{and} \bibinfo{author}{\bibfnamefont{{\v Z}.}~\bibnamefont{{Bo{\v
  s}njak}}}, \bibinfo{journal}{\mnras} \textbf{\bibinfo{volume}{431}},
  \bibinfo{pages}{3550} (\bibinfo{year}{2013}).

\bibitem[{\citenamefont{{Harari} and {Sikivie}}(1992)}]{1992PhLB..289...67H}
\bibinfo{author}{\bibfnamefont{D.}~\bibnamefont{{Harari}}} \bibnamefont{and}
  \bibinfo{author}{\bibfnamefont{P.}~\bibnamefont{{Sikivie}}},
  \bibinfo{journal}{Physics Letters B} \textbf{\bibinfo{volume}{289}},
  \bibinfo{pages}{67} (\bibinfo{year}{1992}).

\bibitem[{\citenamefont{{Fedderke} et~al.}(2019)\citenamefont{{Fedderke},
  {Graham}, and {Rajendran}}}]{2019arXiv190302666F}
\bibinfo{author}{\bibfnamefont{M.~A.} \bibnamefont{{Fedderke}}},
  \bibinfo{author}{\bibfnamefont{P.~W.} \bibnamefont{{Graham}}},
  \bibnamefont{and}
  \bibinfo{author}{\bibfnamefont{S.}~\bibnamefont{{Rajendran}}},
  \bibinfo{journal}{arXiv e-prints}  (\bibinfo{year}{2019}),
  \eprint{1903.02666}.

\end{thebibliography}

\end{document}